\documentstyle[epsfig,12pt]{article}

\begin{document}

\hoffset = -1truecm \voffset = -2truecm \baselineskip = 10 mm

\title{\bf Are valence quarks rotating?}

\author{
{\bf Wei Zhu} \\
\normalsize Department of Physics, East China Normal University,
Shanghai 200062, P.R. China \\ }

\date{}

\newpage

\maketitle

\vskip 3truecm

\begin{abstract}

   We suggest to compare the deep inelastic scattering structure
functions measured in the unpolarized charged-lepton scattering off
a transversely polarized proton and off a longitudinally polarized
proton at larger Bjorken variable $x$, one may get a direct evidence
concerning the valence quark orbital angular momentum in the proton.
This possible rotating effect is estimated by using a
proton-ensemble model.

\end{abstract}


$keywords$: Orbital angular momentum; Two kinds of the EMC effects

\newpage

    The European Muon Collaboration (EMC) result [1] indicates that only a small fraction of the
nucleon spin is due to quark spins. There have been many attempts to
explain this (second) EMC effect from the fundamental theory. The
key to understand the spin structure of the nucleon is the orbital
angular momentum (OAM) of quarks and gluons. However, progress in
this direction has been hindered by a number of difficulties in
measuring.

    In this letter we suggest a direct observation of the valence quark OAM:
comparing the deep inelastic scattering (DIS) structure functions
measured in the unpolarized charged-lepton scattering off a
transversely polarized proton and off a longitudinally polarized
proton at larger Bjorken variable $x$, one may get a direct evidence
concerning the orbital motion of the valence quarks. Our idea is
straightforward: Fermi motion momentum ($\sim 200 MeV$) of a nucleon
inside a nucleus is small comparing with a larger longitudinal
momentum of this nucleon at the infinite momentum frame in DIS,
however, such motion distorts the quark momentum distributions in
the nucleon and it has been clearly observed by so called the
(first) EMC effect [2]. The reason is that the quark distributions
at lager $x$ are small values, any perturbations on them can be
sensitively appeared in the ratio of the structure functions with
and without perturbation. Therefore, we also can expect that the
orbital motion of a valence quark in a transversely polarized proton
will add to the longitudinal motion of this quark and smears its
distribution. On the other hand, there is no such smearing effect in
a longitudinally polarized proton since the rotating plane is
perpendicular to the $z$-direction. Thus, the ratio of the inclusive
structure functions in the above two different polarized protons
will appear a like-EMC effect at larger $x$. We believe that such
evidence has existed in the experimental data base and we have not
used it.

    For estimating the smearing effect of the orbital motion in the valence quark distributions, we construct
a proton-ensemble to analogy Fermi motion in the nucleus.
Considering a DIS event (Fig.1a), where a virtual photon probes a
valence quark in the polarized proton. Because the parton OAM
$\vec{L}$ in a proton is always restricted by a general condition
$L_z+s=1/2$ and it is irrelevant to the independent motions of the
partons, therefore, we assume that the collective rotation of the
partons can be separated from other single particle (intrinsic)
motion at the adiabatic approximation. We denote the velocities of
these two motions at center of mass (c.m) system of the proton are
$\vec{v}_R$ and $\vec{v}_{in}$, respectively. Now we image that this
targeted proton $P$ is replaced by an equivalent non-rotating proton
$P^*$, which moves with velocity $\vec {v}_R$ but keeps all other
non-rotating structures of the proton $P$ (Fig.1b). The probe can
not distinguish two scattering events in Fig. 1a and 1b, however, we
realize the separation of the orbital rotating motion from other
intrinsic motions of the struck quark and avoid the unknown
parameter-the mass of the current quarks. After A-times
measurements, we have a proton-ensemble $\{P^*\}_A$, which is
consisted by A-protons $P^*$(Fig. 1b). The normalized distribution
of the equivalent proton $P^*$ inside the proton-ensemble
$\{P^*\}_A$ at its c.m system is $\rho_{\xi}(\vec{k}_R)$, where
$\xi$ indicates the polarized state of the original proton and
$\vec{k}_R=m_N\vec{v}_R$. We project this distribution to the
plus-component in the light-front version, one can get the
probability of an equivalent proton $P^*$ caring the fractional
momentum $y=Ak^+_R/P^+_{\{P^*\}_A}$

$$f_{R,\xi}(y)=\int d^3\vec{k}_R\rho_{\xi}(\vec{k_R})\delta(y-Ak^+_R/P^+_{\{P^*\}_A}). \eqno(1)
$$

    We denote the quark momentum distribution in the equivalent proton
$P^*$ as $F_q(z,Q^2)$ and that in the proton-ensemble as
$F_{q,\xi}(x,Q^2)$, they have relation

$$F_{q,\xi}(x,Q^2)=\int dyf_{R,\xi}(y)F_q(x/y,Q^2),\eqno(2)$$ where

$$x=\frac{Ak^+_{in}}{P^+_{\{P^*\}_A}};~~~z=\frac{x}{y}=\frac{k^+_q}{k^+_R}.\eqno(3)$$

    For the longitudinally polarized proton with respect to its motion, situation becomes simple.
Since

$$\rho_{L}(\vec{k}_R)=\rho_{L}(k_{R,x},k_{R,y})\delta(k_{R,z}),\eqno(4)$$ we
have

$$f_{R,L}(y)=\delta(1-y).\eqno(5)$$

    On the other hand, for the transversely-polarized proton,

$$\rho_{T}(\vec{k}_R)=\rho_{T}(k_{R,y},k_{R,z})\delta(k_{R,x}).\eqno(6)$$

    We take $A$ as a large number. Since the rotating momentum is small comparing with a
larger longitudinal momentum in the infinite momentum frame, its
distribution from $0$ to $k_{R,max}$ is narrow. For simplicity we
assume that the distribution $\rho_{T}(k_{R,y},k_{R,z})$ presents
roughly a uniform distribution from $\sqrt{k^2_{R,y}+k^2_{R,z}}=0$
to $k_{R,max}$, i.e.,

$$\rho_{T}(k_{R,y},k_{R,z})=\frac{1}{\pi
k^2_{R,max}}\theta\left(k_{R,max}-\sqrt{k^2_{R,y}+k^2_{R,z}}\right).\eqno(7)$$
Thus, we have

$$f_{R,T}(y)=\left\{\begin{array}{ll}\frac{2}{\pi}\left(\frac{m_N}{k_{R,max}}\right)^2\sqrt{\left(\frac{k_{R,max}}
{m_N}\right)^2-(y-1)^2} &{\rm if}~ 1-k_{R,max}/m_N\le y\le1+k_{R,max}/m_N\\
0 &{\rm otherways}\end{array}\right. \eqno(8)$$

    We use the ratio

$$R_I\stackrel{x>0.4}{=}\frac{F_{2,T}(x,Q^2)}{F_{2,L}(x,Q^2)}\eqno(9)$$ to
describe the rotating effect of the valence quarks, where
$F_{2,T}(x,Q^2)$ (or $F_{2,L}(x,Q^2)$) is the structure function of
the unpolarized lepton scattering off a transversely polarized
proton (or off a longitudinally polarized proton). A convenient
method is to measure

$$R_{II}\stackrel{x>0.4}{=}\frac{F_2(x,Q^2)}{F_{2,L}(x,Q^2)},\eqno(10)$$
where $F_2(x,Q^2)$ is the structure function of the unpolarized
proton, although the expected effect in $R_{II}$ is about 2/3 of
$R_{I}$.

    For estimating the rotating effect, we
use a simple parametrization of the valence quark distribution
$xV(x,Q^2)=3B^{-1}(0.5,4)x^{0.5}(1-x)^{3}$, where B is the Beta
function. In Fig. 2 we present the results using $k_{R,max}=20, 50,
100$ and $200 MeV$, which corresponding to the maximum rotating
velocities $v_{R,max}=0.02, 0.05, 0.11$, and $0.21$ in the natural
unit $\hbar=c=1$. Following a consideration in classical mechanics,
we regard the target proton as a rigid body rotating with angular
velocity $\omega=v_{R,max}/R_N$. Thus we can estimate the
contributions of the parton OAM to the proton spin in the above
mentioned examples, they are $L/(1/2\hbar)=I\omega/(1/2\hbar)=8\%,
20\%, 40\%$ and $80\%$, respectively. Although these predictions are
naive and we do not expect the precise OAM-information from them,
however, any results deviating (even non-deviating) from unity in
Eq. (9) are instructive for understating "spin crisis".

    In conclusion, through the relations between two kinds of EMC effects
(i.e., the contributions of Fermi motion to the first EMC effect and
the spin structure of the proton in the second EMC effect), we
suggest to compare the deep inelastic scattering structure functions
measured in the unpolarized charged-lepton scattering off a
transversely polarized proton and off a longitudinally polarized
proton at larger Bjorken variable $x$, one may get a direct evidence
concerning the valence quark orbital motion. This possible rotating
effect is estimated by using a proton-ensemble model.

\newpage

\newpage
\vskip 1.0 truecm \hbox{
\centerline{\epsfig{file=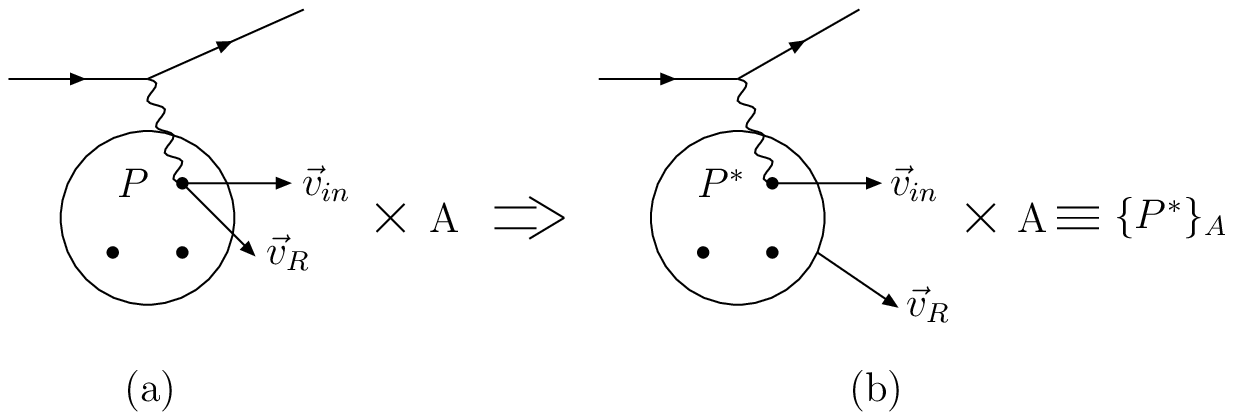,width=15.0cm,clip=}}} Fig. 1 (a) A
DIS event of the unpolarized charged-lepton scattering off a
polarized proton $P$, where the struck quark has rotating velocity
$\vec{v}_R$ and other intrinsic velocity $\vec{v}_{in}$ ; (b) An
equivalent DIS event corresponding to (a), where target proton $P^*$
moves with $\vec{v}_R$ without rotation; A-times measurements
construct a proton-ensemble $\{P^*\}_A$. \vskip 1.0 truecm

\newpage
\vskip 1.0 truecm \hbox{
\centerline{\epsfig{file=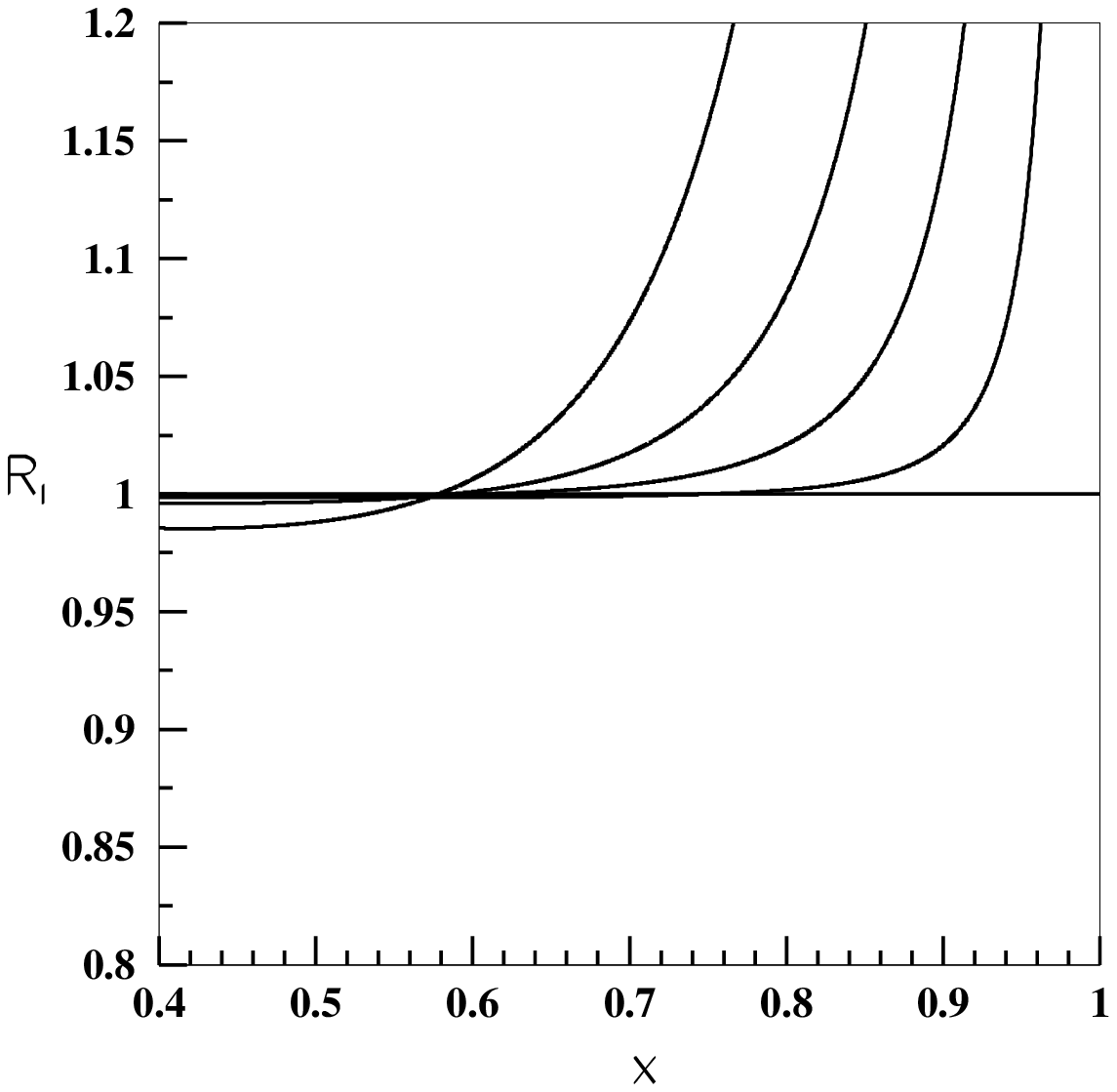,width=15.0cm,clip=}}} Fig. 2 The
ratio $R_I$ in Eq. (9), which presents the smearing effect of
different orbital angular momenta of the valence quarks with
parameter (from right to left) $k_{R,max}=20, 50, 100$ and $200
MeV$. \vskip 1.0 truecm

\end{document}